\documentclass[twocolumn,showpacs,prb,amsmath,amssymb,superscriptaddress]{revtex4}
\usepackage{graphicx}
\usepackage{dcolumn}
\usepackage{bm}
\usepackage{mathrsfs}

\def\ke#1{|#1 \rangle }

\def\avof#1#2#3{\langle \,#1 \,|\, #2 \,|\,#3\,\rangle }

\begin{document}

\title{Spin Hall Effect of Excitons}

\author{Shun-ichi Kuga}
 \affiliation{Department of Applied Physics, University of Tokyo, \\
 7-3-1 Hongo, Bunkyo-ku, Tokyo 113-8656, Japan}
\author{Shuichi Murakami}
\thanks{Corresponding Author}
 \email{murakami@stat.phys.titech.ac.jp}
 \affiliation{Department of Physics, Tokyo Institute of Technology, \\
2-12-1 Ookayama, Meguro-ku, Tokyo 152-8551, Japan}
\affiliation{PRESTO, Japan Science and Technology Corporation (JST), Saitama, 332-0012, 
Japan}\author{Naoto Nagaosa}
 \affiliation{Department of Applied Physics, University of Tokyo, \\
 7-3-1 Hongo, Bunkyo-ku, Tokyo 113-8656, Japan}
\affiliation{Cross-Correlated Materials Research Group (CMRG), ASI, RIKEN, Wako 351-0198, Japan}

\begin{abstract}
Spin Hall effect for excitons in alkali halides and in Cu$_2$O 
is investigated
theoretically. 
In both systems, the spin Hall effect results from the Berry curvature 
in $k$ space, which becomes nonzero due to 
lifting of degeneracies of the exciton states by exchange coupling. 
The trajectory of the excitons can be directly seen as
spatial dependence of the circularly polarized light emitted from 
the excitons. It enables us to observe the spin Hall effect 
directly in the real space-time. 
\end{abstract}

\maketitle

\section{Introduction}
Spin Hall effect(SHE) is attracting interest recently 
because it can produce spin 
current without magnetism or magnetic field. 
The research was triggered by the two theoretical proposals on the 
intrinsic mechanism on the SHE \cite{Mur,Sin}, 
and it has been intensively studied both theoretically and experimentally. 
There are various experiments on the SHE in doped semiconductors 
and in metals\cite{Kato,Wun,Saitoh,Valenzuela} by optical 
and electrical methods. 
In these observations in electronic systems, the spin
current is seen as an effect summed over many electrons, 
while the motion of the
 individual electrons cannot be seen. Therefore,
comparison between theory 
and experiments is sometimes indirect and not straightforward. 
An experimental method to see directly the electron trajectory is highly 
desired. At first sight 
it seems impossible because  
condensed materials have
a huge number of electrons, which 
cannot be distinguished from each other.

Apart from electronic systems, we have one example where 
one can observe directly the SHE as a  
trajectory of the particle: light \cite{M.Onoda1}. 
As the intrinsic SHE is induced by the Berry phase, it is not limited to electronic systems
but also seen in other (even classical) wave phenomena such as light. 
In this SHE of light, the difference of the refractive indices at an interface of two different media plays the role of 
the
``electric field" in the electronic SHE. 
The SHE of light at the interface is recently measured in 
a high accuracy of about 1$\mathrm{\AA}$ using weak measurement\cite{Hosten}. 

In this letter we theoretically propose a 
way to optically observe 
the trajectory of an elementary excitation driven by the SHE.
We consider two candidates; 
transverse 
excitons in alkali halides and orthoexcitons in Cu$_{2}$O.
We propose an experimental setup, 
and estimate the shift size due to the SHE, which turns out to be
enough magnitude for observation.
In both systems, an electron-hole exchange coupling lifts the degeneracy 
of the excitonic states,
which gives rise to the Berry curvature in $k$ space of the center-of-mass motion. It leads to the 
SHE, namely spin-dependent trajectory of the excitons.
After the radiative lifetime,
these excitons emit light,
whose circular polarization is determined by 
the exciton spins.
Thus by spatially resolving the circular polarization of the emitted light, we can see how the excitons move in real space in a 
spin-dependent way. 
It is the first proposal of a real-space observation 
of the Berry-phase-driven
SHE in electronic systems.

\section{Spin Hall Effect of excitons in alkali halides}
Due to the spin-orbit coupling, 
exciton states in alkali halides with the lowest energy 
consists of an electron in the $\Gamma_{6}^{+}$ conduction band,
and a hole in the 
$\Gamma_{8}^{-}$ valence band,
and these states
are further classified into pure spin-triplet states 
(total angular momentum $J=2$)
and spin singlet-triplet mixed states ($J=1$). Exchange
interaction and the spin-orbit coupling lifts the degeneracy
among these states \cite{Ono}, and
the energies of the $J=2$ excitons are lower than those of the $J=1$
due to the analytic exchange interaction. 
The $J=1$ excitons are allowed for optical dipolar transition,
and are suitable for real-space imaging of the SHE.
Meanwhile, the $J=2$ states 
are dipolar forbidden. 
Hence we restrict ourselves to the $J=1$ excitons. 
The nonanalytic 
exchange Hamiltonian with the basis 
$\{\ke{O_{x}},\ke{O_{y}},\ke{O_{z}}\}$ within the $J=1$ states
is given by \cite{Cho}
\begin{equation}
H_{ex}(\vec{K})=
\frac{\Delta_{\mathrm{LT}}}{K^{2}}(K^{2}-
(\vec{K}\cdot \vec{S})^{2}),
\end{equation}
where $\vec{S}$ is the set of the spin-1 matrices.
$\Delta_{\mathrm{LT}}$ is the longitudinal-transverse (L-T) 
splitting, which can be experimentally determined e.g. from
 polarization beating of the emission \cite{Langbein07}.
We neglect higher order terms in $\vec{K}$.
In addition, for simplicity, we assume that the analytic exchange 
(the splitting between $J=1$ and $J=2$) is 
much larger than the nonanalytic one $\Delta_{\mathrm{LT}}$.
In the calculation of the Berry curvature, 
this assumption allows us to retain only the 
matrix elements within the $J=1$ states among 
the various matrix elements in the 8$\times$8 Hamiltonian 
in the space spanned by $J=1$ and $J=2$ states (see Ref.~\onlinecite{Cho} and 
Table 8 in Ref.~\onlinecite{Hoenerlage}).
This Hamiltonian $H_{ex}$ 
is diagonalized by eigenstates of the helicity
$\lambda=(\vec{K}\cdot \vec{S})/K$ 
with eigenvalues of $\lambda=\pm 1,0$.
Hence, the eigenstates of $H_{ex}(\vec{K})$ are 
 twofold degenerate transverse modes 
and a longitudinal mode, 
whose energies differ by $\Delta_{\mathrm{LT}}$.
This L-T splitting gives rise to the Berry curvature for the $J=1$ excitons,
leading to the SHE.

When the eigenstates are degenerate, a
 wavepacket  follows the semiclassical 
equations of motion\cite{Mur,Sundaram,Shindo,Niu2}:
\begin{eqnarray}
&&\dot{\vec{R}}_{c}=\frac{1}{\hbar }\frac{\partial \varepsilon_{n}(\vec{K}_{c})}{\partial \vec{K}_{c}}+\dot{\vec{K}}_{c}\times \eta^{\dag}\vec{{\cal F}}_{n}(\vec{K}_{c})\eta, \label{EOM-r}\\
&&\hbar \dot{\vec{K}}_{c}=-\frac{\partial V(\vec{R}_{c})}{\partial \vec{R}_{c}}, \ \ \ \   \dot{\eta}=-i\dot{\vec{K}}_{c}\cdot \vec{{\cal A}}_{n}(\vec{K}_{c}) \ \eta,  
\label{EOM}
\end{eqnarray}
where $\vec{R}_{c},\vec{K}_{c}$ are the center position and the wavevector of the wavepacket, $\varepsilon_{n}(\vec{K}_{c})$ is the energy dispersion
of the $n$-th band, $V(\vec{R}_{c})$ is an external potential, and $\eta=(\eta_{1},\eta_{2})$ is the internal degree of freedom of the two degenerate transverse exciton  bands.
$\vec{{\cal A}}_{n}(\vec{K})$ and $\vec{{\cal F}}_{n}(\vec{K}_{c})$ are Berry connection and Berry curvature, which are defined as
\begin{eqnarray}
\left\{
\begin{array}{l}
\displaystyle [{\cal A}_{n}^{\mu}(\vec{K})]_{ij}\equiv -i\avof{n_{i}(\vec{K})}{\frac{\partial}{\partial K_{\mu}}}{n_{j}(\vec{K})}, \\
\displaystyle {\cal F}_{n}^{\rho}(\vec{K})\equiv \epsilon_{\mu\nu\rho}\left(
\frac{\partial {\cal A}_{n}^{\nu }(\vec{K})}{\partial K_{\mu}}+i{\cal A}_{n}^{\mu}(\vec{K}){\cal A}_{n}^{\nu}(\vec{K})\right), 
\end{array}
\right. \label{BC}
\end{eqnarray}
where $\ke{n_{i}(\vec{K})}$ is an eigenstate of the 
$n$-th band and $i$ is the label for each eigenstate within the 
degenerate band.  
The term $\dot{\vec{K}}\times \eta^{\dag}\vec{{\cal F}}\eta $ in 
the equation of motion for $\dot{R}_{c}$ 
is called  anomalous velocity, which leads to the SHE. 
The Berry phase changes sign when the spin direction is reversed.
Therefore, two wavepackets with opposite spins move along 
opposite directions to each other.   
This mechanism is responsible for the SHE of electrons in p-type 
semiconductors \cite{Mur}
and that of light\cite{M.Onoda1}.

The Berry curvature for the $J=1$ exciton states can be calculated 
from $H_{ex}$   
in the same way
as that in the SHE of light \cite{M.Onoda1}, 
because the two cases share the same feature of L-T splitting in the 
spin-1 systems. Therefore the Berry curvature of 
the transverse states with helicity of $\lambda=\pm 1$
is then calculated as 
\begin{equation}
{\cal F}_{n}^{\rho}(\vec{K})=\lambda \frac{K_{\rho}}{K^{3}}.
\end{equation}
The longitudinal state ($\lambda=0$) has a vanishing Berry curvature, and 
it does not undergo a shift due to the SHE.

\begin{figure}[htbp]
  \begin{center}
    \includegraphics[width=8cm]{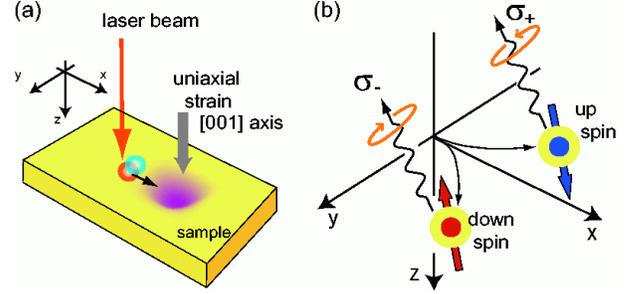}
  \end{center}
  \caption{(a) Experimental setup for the detection of SHE in alkali halides. The twofold degenerate wavepacket of the transverse excitons moves toward the center of the uniaxial trapping potential, along the $x$ direction. 
(b) Schematic figure of the spin Hall effect of excitons. The up- and
down-spin 
wavepackets are deflected to the  $\mp y$ directions, respectively,
and they 
emit 
light with opposite circular polarizations. }
  \label{fig1}
\end{figure}

We propose an experiment
to detect the SHE in the real space and 
evaluate the Hall shift. 
The SHE requires a nonzero $\dot{\vec{K}}_{c}$ as seen 
from Eq.~(\ref{EOM}). 
Namely, one should apply an external force to the exciton to 
see a shift due to the SHE. For electrons an electric field is sufficient,
whereas an exciton cannot 
be accelerated by an electric field.
Instead, a local strain gives rise to a potential gradient and accelerates
excitons, inducing the SHE. 
Thus we propose the following setup;
we prepare an 
transverse exciton wavepacket with momentum along the 
$z$ direction, and apply a uniaxial local strain,
so that the excitons feel a force along the $x$ 
direction, as shown in Fig.\ref{fig1}.

A strain-induced potential well has been 
developed for Cu$_2$O \cite{Naka}, 
but not for alkali halides to our knowledge. 
Therefore, we estimate the shift from existing data on alkali halides.
From the data on the thin-film RbI for example, the effect of 
uniaxial strain is 25-45 meV for 1 kbar \cite{Ohno,Nishimura,Itoh}.
Because the crystal is easily broken by
high uniaxial pressure, we take a lower value for a trapping potential,  
4meV for 0.1kbar as an example.
We assume the size of the trap to be several hundred micrometers, 
as developed for Cu$_2$O \cite{Naka}.
Thus we consider a 200$\mathrm{\mu m}$-4meV configuration of the trapping potential.
The force acting on the exciton wavepacket is $3.2\times 10^{-18}$N,
and the corresponding rate of the wavevector change is 
$\dot{K}_{x}\simeq 3.0\times 10^{16}\mathrm{m^{-1}s^{-1}}$. 
When we take RbI for example, the typical wavenumber is $k_0=0.8\times 
10^6$cm$^{-1}$.
The magnitude of the Berry curvature is
$F^{z}= k_{0}^{-2}\simeq 1.6\times 10^{-16}\mathrm{m^{2}}$.
Therefore the anomalous velocity is 
$v_{\mathrm{a}}=\dot{K}_{x}F^{z}\simeq 4.8$m/s and the shift is 
$y_{\mathrm{a}}=v_{\mathrm{a}}\tau\simeq 8\mathrm{n m}$, 
where $\tau=1.7$ns is the lifetime of the exciton in RbI,
which is governed by self-trapping process \cite{Tsu}. 
We note that this self-trapping instability can be reduced or avoided by choosing other materials such as III-V or II-VI compounds, AgBr, and TlBr, where the free state of exciton is more stable than the self-trapped state. 
In these materials, the shift of the excitons could be much 
longer \cite{toyozawa}.

Because of the uncertainty principle, 
in order for the wavepacket to have a well-defined wavenumber, the
size of the wavepacket in $k$ space
should be much larger than the wavenumber. 
Hence the ratio between the size of the exciton wavepacket 
and the transverse shift is small, and 
the direct observation of the SHE might be difficult.
Nevertheless, a wavepacket deflected to the transverse direction is 
spin-polarized and emits a circularly polarized light.
Therefore, one can observe the SHE by detecting the spatial dependence
of the circular polarization from the two wavepackets deflected in the opposite
direction.

\section{Spin Hall Effect of orthoexciton in Cu$_{2}$O}
In Cu$_{2}$O,
the exciton states with the lowest energy, composed of the 
$\Gamma_{7}^{+}$-valence band and the conduction band, is the $1S$ exciton.
Because the valence band and the conduction
band share the same parity, 
radiative recombination of the $1S$ exciton is dipolar forbidden, and 
therefore this state has a long radiative 
lifetime.  The four states in the $1S$ yellow excitons are classified into 
three $\Gamma_{5}^{+}$  
orthoexciton states 
and one $\Gamma_{2}^{+}$ paraexciton state.
The orthoexcitons are singlet-triplet mixed states, while the 
paraexciton is purely spin-triplet. Therefore exchange
interaction exists only in the singlet states, and 
affects only the energy of the orthoexcitons, while the paraexcitons
remain intact.
The energy splitting between 
ortho and paraexcitons due to the exchange interaction is about 12meV.
Furthermore, 
the degeneracy of the three orthoexciton states
is lifted by (nonanalytic)  exchange splitting.
The matrix form of the exchange interaction among the orthoexciton states
$\{\ke{O_{yz}},\ke{O_{zx}},\ke{O_{xy}}\}$ is given as
\begin{widetext}
\begin{eqnarray}
H_{ex}(\vec{K})=\left[
  \begin{array}{ccc}
    \Delta_{Q}\frac{K_{y}^{2}K_{z}^{2}}{K^{2}}+\Delta_{3}(3K_{x}^{2}-K^{2})   & (\Delta_{Q}\frac{K_{z}^{2}}{K^{2}}+\Delta_{5})K_{x}K_{y}  &  (\Delta_{Q}\frac{K_{y}^{2}}{K^{2}}+\Delta_{5})K_{z}K_{x}  \\
    (\Delta_{Q}\frac{K_{z}^{2}}{K^{2}}+\Delta_{5})K_{x}K_{y}   &  \Delta_{Q}\frac{K_{z}^{2}K_{x}^{2}}{K^{2}}+\Delta_{3}(3K_{y}^{2}-K^{2})  &  (\Delta_{Q}\frac{K_{x}^{2}}{K^{2}}+\Delta_{5})K_{y}K_{z}  \\
    (\Delta_{Q}\frac{K_{y}^{2}}{K^{2}}+\Delta_{5})K_{z}K_{x}   &  (\Delta_{Q}\frac{K_{x}^{2}}{K^{2}}+\Delta_{5})K_{y}K_{z}  &  \Delta_{Q}\frac{K_{x}^{2}K_{y}^{2}}{K^{2}}+\Delta_{3}(3K_{z}^{2}-K^{2})  \\
  \end{array}
\right]. \label{ex-c}
\end{eqnarray}
\end{widetext}
where the parameters are $\Delta_{Q}k_{0}^{2}=5.0\mathrm{\mu eV},\Delta_{3}k_{0}^{2}=-1.3\mathrm{\mu eV},\Delta_{5}k_{0}^{2}=2.0\mathrm{\mu eV} $\cite{dasbach} with the wavenumber $k_{0}\equiv 2.62\times 
10^{7}\mathrm{m^{-1}}$, as obtained experimentally from 
the high resolution spectroscopy of polaritons \cite{dasbach}.

\begin{figure}[htbp]
  \begin{center}
    \includegraphics[width=7.5cm]{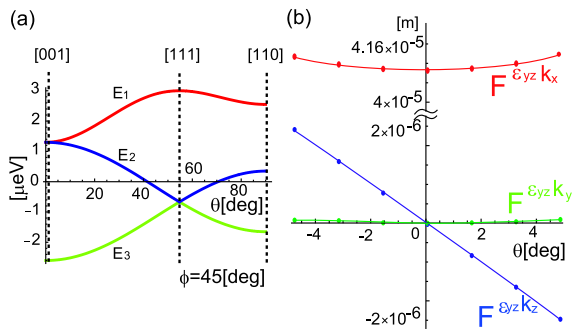}
  \end{center}
  \caption{(a) Energy dispersion of the exchange interaction 
and (b)distribution 
of the Berry curvature $F_{n}^{\epsilon\mu}$ in Cu$_2$O. They are shown  
as a function of the polar angle $\theta $ of $\vec{K}$, with the 
azimuthal angle $\phi=45^{\circ}$. 
The strain $\epsilon_{yz}$ is set to be zero. }
  \label{fig2}
\end{figure}
The wave-vector dependence of the exchange interaction (\ref{ex-c}) is illustrated in Fig.~\ref{fig2}(a).
The eigen energies $E_{1}(\vec{K})$ and $E_{2}(\vec{K})$ are degenerate along the $[0\,0\,1]$ direction and $E_{2}(\vec{K})$ and $E_{3}(\vec{K})$ are degenerate along the $[1\,1\,1]$ direction.
One possible experiment is to make a potential trap exert a 
force to the exciton, as we considered in alkali 
halides.
In Cu$_2$O, however, the strain is 
typically of the order of meV, much larger than 
the exchange coupling ($\sim\mu$eV).  Hence one cannot
ignore the strain in the calculation of the
Berry curvature. 
This local strain 
in general reduces considerably the Berry curvature stemming from the exchange 
coupling, because of its larger energy scale. 
To overcome this difficulty, we consider another type of 
strain: a shear strain $\epsilon\equiv\epsilon_{yz}$. 
The shear strain brings about an additional term to the Hamiltonian 
as
$H'_{ij}=
\Lambda\epsilon_{yz}(\delta_{i2}\delta_{j3}
+\delta_{i3}\delta_{j2})$. 
For simplicity we change the normalization of 
the dimensionless strain parameter $\epsilon\equiv \epsilon_{yz}$,
by taking $\Lambda=8.1\textrm{meV}$
which is the energy shift expected for 5kbar shear strain,
that is calculated from the data in \cite{Naka}.
Using this we consider a Berry curvature in the hyperspace of 
$\epsilon$-$\vec{K}$, which follows \cite{Sundaram}
\begin{equation}
\dot{R}_{\mu}=\frac{1}{\hbar}
\frac{\partial E_{n}}{\partial K_{\mu}}
-\dot{\epsilon}
\eta^{\dag}
{\cal F}_{n}^{\epsilon\mu}(\vec{K})\eta. 
\end{equation}
with Berry connection and Berry curvature that are defined as
\begin{align}
& [{\cal A}_{n}^{\epsilon}(\vec{K})]_{ij}\equiv -i\avof{n_{i}(\vec{K})}{\frac{\partial}{\partial \epsilon}}{n_{j}(\vec{K})}, \\
& {\cal F}_{n}^{\epsilon\mu}(\vec{K})\equiv 
\frac{\partial {\cal A}_{n}^{\mu }}{\partial \epsilon}
-\frac{\partial {\cal A}_{n}^{\epsilon }}{\partial K_{\mu}}
+i\left[{\cal A}_{n}^{\epsilon},\ {\cal A}_{n}^{\mu}\right], 
\end{align}

Because
the Hamiltonian matrix $H_{ex}+H'$ is real, the eigenvectors 
can be chosen as real. 
The Berry curvature ${\cal F}^{\epsilon\mu}(\vec{K})$ is 
then pure imaginary and Hermitian.
If the state considered is nondegenerate, the Berry curvature 
is scalar ($1\times 1$ matrix), and therefore it vanishes. 
On the other hand, when the state is twofold degenerate, as in [001] 
or in [111] direction, 
the  Berry curvature is a 2$\times$2 matrix. 
It is therefore proportional to the 
Pauli matrix $\sigma_{y}$:
\begin{equation}
{\cal F}^{\epsilon\mu}(\vec{K})=F^{\epsilon\mu}(\vec{K})\sigma_{y}.
\end{equation}
Thus to see the SHE, the exciton states should be degenerate, which 
occurs along the high-symmetry lines.
For concreteness, we hereafter focus on the twofold 
degeneracy along the $[0,0,1]$ direction ($\vec{K}\|\hat{z}$) as the degenerate bands in the semiclassical equation of motion
(\ref{EOM}). Then the eigenstates $\ke{n_{1}(\vec{K})}$ and $\ke{n_{2}(\vec{K})}$ with eigenenergies $E_{1}(\vec{K})$ and $E_{2}(\vec{K})$ in Fig.~\ref{fig2}(a) are considered as  pseudospin states. 
Along the $[0\,0\,1]$ direction, these states become $|O_{yz}\rangle$ 
and $|O_{zx}\rangle$.

Figure \ref{fig2}(b) is the distribution of 
$F^{\epsilon\mu}$. 
We note that $F^{\epsilon\mu}$ depends on gauge, even though the anomalous velocity does not, and 
Fig.~\ref{fig2}(b) is based on a particular gauge choice.
The typical size of the Berry curvature is expected to be
$F\sim(\Lambda/\Delta_{\mathrm{gap}})k_{0}$ from consideration  of 
relevant energy scales,
where $\Delta_{\mathrm{gap}}$ denotes the energy gap between 
the ($\ke{n_1}$, $\ke{n_2}$) states and the $\ke{n_3}$ state. 
Because $\Lambda$ and $\Delta_{\mathrm{gap}}$ are
of the order of meV and $\mu$eV, respectively, 
this estimate agrees with   
Fig.~\ref{fig2}(b). 

In fact, for $\vec{K}\|\hat{z}$ the Berry curvature can 
be calculated analytically as ${\cal F}^{\epsilon x}(\vec{K})=
(\Delta_5\Lambda)/(9\Delta_3 k_0^3)=4.06\times 10^{-5}\mathrm{m}$, and the other components
are zero:
${\cal F}^{\epsilon y}(\vec{K})=0$, ${\cal F}^{\epsilon z}(\vec{K})=0$.
The reason for the vanishing $y$ and $z$ components is the mirror symmetry
with respect to the $yz$ plane, and the
twofold rotational symmetry around the $z$ axis, respectively. 
Therefore, for $\vec{K}$ along the $[0\,0\,1]$ direction,  the 
anomalous velocity is along the $x$ 
direction.
Because the SU(2) Berry curvature ${\cal F}^{\mu}(\vec{K})$
is proportional to $\sigma_y$, 
we take the eigenvectors of $\sigma_{y}$, i.e.
$\frac{1}{\sqrt{2}}\binom{1}{\pm i}$
(in the $\ke{n_{1}}$-$\ke{n_{2}}$ basis),
and the semiclassical
equations of motion (\ref{EOM}) is diagonalized. 
In this basis, the spin $\eta$ only acquires U(1) phase
in time evolution, but does not change its direction. 
Therefore, for the wavenumber along the $[0\,0\,1]$ direction, 
the wavepackets for $(\ke{O_{xz}}\pm i\ke{O_{yz}})/\sqrt{2}$ 
have opposite anomalous velocity, and their spins are along $\pm z$, 
respectively.
These excitons emit circularly polarized light depending
on its spin state \cite{elliott}.
This enables us to see this spin Hall shift directly by 
an optical method.

The anomalous velocity is proportional to $\dot{\epsilon}$. 
Therefore, 
in order to induce the SHE, the strain should be varied 
externally. One may consider adding an oscillating strain with 
frequency $\omega$. 
Then the typical size of 
the shift is 
$\epsilon(\Lambda/\Delta_{\mathrm{gap}})/k_{0}\sim
(E_{\mathrm{str.}}/\Delta_{\mathrm{gap}})/k_{0}$, where
$E_{\mathrm{str.}}(\sim\Lambda\epsilon)$ is the  energy 
shift of excitons by strain.
Thus only the small strain of the order of $\mu$eV gives rise 
to the shift of the order of a wavenumber $\sim 600$nm. 
Although the radiative lifetime is $\tau_{\mathrm{rad.}}\sim 14
\mu\mathrm{s}$ \cite{Ohara}, the lifetime of the orthoexcitons is much shorter:
$\tau\sim 3 \mathrm{ns}$, due to a nonradiative rapid conversion 
from orthoexcitons to paraexcitons. 
The oscillation of the strain $\epsilon$ should be faster
than $1/\tau$, i.e. be as fast as gigahertz in frequency.

The light emission from the orthoexciton may be reduced by 
several reasons.
First, among all the orthoexcitons only the fraction of
$\tau/\tau_{\mathrm{rad.}}\sim 2\times 10^{-4}$ emit light. 
The resolution to
detect this emission is it to be 
well achievable, 
because the radiative decay
rate of excitons has been measured in experiments \cite{Ohara}. 
Furthermore, when the density of the orthoexcitons exceed a critical 
value ($\sim 10^{15}\mathrm{cm}^{-3}$), 
the spin exchange process between two 
orthoexcitons will be effectively convert orthoexcitons
into paraexcitons in a short timescale ($\sim 100\mathrm{ps}$)
 \cite{Kubouchi}. 
A typical density of
excitons by continuous wave (CW) 
laser is $10^{13}$-$10^{14}\mathrm{cm}^{-3}$; it is 
well below the critical density, and it is not a problem for 
the proposed experiment.
The interaction between orthoexcitons also leads to phase decoherence,
but it does not affect the SHE, as Eqs.\ (\ref{EOM-r})(\ref{EOM})
remains unaffected. 
 This situation is similar to the electrons in solids, 
 where the mean free path is much shorter than the excitons, but still 
 shows the spin Hall effect. This is because the spin Hall effect is the 
 accumulative effect of the transverse motion of the particles, which does
 not require the coherence of the process. 

\section{Summary and discussions}
In conclusion, we theoretically investigate the SHE of the excitons in alikali halides and in Cu$_{2}$O. The exchange coupling lifts the threefold 
degeneracy of the orthoexcitons, while in some directions of the wavenumber 
double degeneracy remains. This remaining double degeneracy gives rise to 
nonzero SU(2) Berry curvature, leading to the SHE. 
This SHE can be observed as a position-dependent
circularly polarized light emitted from the orthoexcitons.

Recently Yao and Niu \cite{Niu1} proposed SHE for excitons in GaAs
quantum well. In their paper the main contribution to
the Berry curvature comes from the heavy-hole light-hole mixing 
in the quantum well, whereas in the present paper the exchange coupling
between the hole and electron spins is 
the main source of the Berry curvature.
Because of the degeneracy of the energy spectrum, the Berry curvature 
is enhanced in our setup, thereby the SHE becomes prominent.
Furthermore, we propose in this paper a realistic setup with target
material specified. 
The proposed setup enables us to use modulation spectroscopy with
high precision. This provides us with a space-time resolved
measurement of the SHE.

As a closely related subject, an optical SHE has been observed in 
an exciton-polariton system \cite{Leyder}, whose mechanism 
is totally different from the SHE in the present paper.
The two different mechanisms for the intrinsic SHE are
(A) precession due to
the $\mathbf{k}$-dependent (Zeeman-like) field acting on the spin,
and (B) the anomalous velocity from the 
$\mathbf{k}$-space Berry curvature. Although
they are often confused with each other, they are distinct.
The mechanism (A) is used 
in the optical SHE in exciton-polaritons  
\cite{Leyder,Kavokin05,Glazov07,Glazov08}, and in the 
SHE in the Rashba system \cite{Sin}. In these cases the spin-orbit coupling
is linear in terms of the spin, which means that the spin-orbit splitting 
can be regarded as a ``Zeeman-like" field, although the external 
magnetic field is zero.
In these systems the mechanism (B) is absent because the 
contribution of the Berry curvature cancels between the two bands
involved. 
On the other hand, the mechanism (B)
causes the SHE in excitons in the present paper,
as well as the SHE in 
the Luttinger model \cite{Mur}. This mechanism works even when
the Hamiltonian is not linear in the spin operator. 
This effect due to (B) 
is enhanced when 
band crossings exist near the Fermi energy, e.g.\  
in the SHE in platinum \cite{Guo}, while it is not the case in (A).
Moreover, (B) gives an additional spin-dependent (anomalous) 
velocity and deflects the exciton trajectory,
while (A) does not. 
Thus the mechanisms (A) and (B) are distinct, and 
the experiments proposed in the present paper allows us a first 
real-space observation of the Berry-curvature mechanism (B) 
in electronic systems.

\begin{acknowledgments}
We are grateful to K. Yoshioka and M. Kuwata-Gonokami for fruitful discussions.
This research is partly supported  
by Grant-in-Aids under the grant numbers 16076205, 17105002, 19019004, 19048008, 19048015, and 19740177
from the Ministry of Education,
Culture, Sports, Science and Technology of Japan.  
\end{acknowledgments}

\end{document}